\documentclass[%
 reprint,
 amsmath,amssymb,
 aps,
 prl,
floatfix,
]{revtex4-1}

\usepackage{graphicx}
\usepackage{setspace}
\usepackage{dcolumn}
\usepackage{minibox}
\usepackage{bm}
\usepackage{hyperref}
\usepackage{float}
\floatstyle{boxed}

\begin{document}

\title{Integrated Information-induced quantum collapse}

\author{Kobi Kremnizer$^{1}$ and Andr$\acute{e}$ Ranchin$^{2,3}$ }

\affiliation{ $^{1}$University of Oxford, Mathematical Institute}
\affiliation{$^{2}$University of Oxford, Department of Computer Science, Quantum Group}
\affiliation{$^{3}$Imperial College London, Department of Physics, Controlled Quantum Dynamics}

\date{\today}

\begin{abstract}

We present a novel spontaneous collapse model where size is no longer the property of a physical system which determines its rate of collapse. Instead, we argue that the rate of spontaneous localization should depend on a system's quantum Integrated Information (QII), a novel physical property which describes a system's capacity to act like a quantum observer. We introduce quantum Integrated Information, present our QII collapse model and briefly explain how it may be experimentally tested against quantum theory.   

\end{abstract}

\maketitle

\section{Introduction}

Quantum mechanics is plagued with latent fundamental issues. The theory admits the delicate coexistence of two radically different dynamics. Unobserved systems undergo linear, deterministic, unitary evolution whereas observation causes a non-linear, probabilistic, non-unitary ``collapse'' of the quantum state. From the outset, the ontological significance of the quantum state is unclear. Moreover, the quantum superposition of distinguishable states and the arising of probabilities seem to contradict the behavior we observe in macroscopic systems. Is there a classical/quantum divide and if so, where does it lay? 

These issues are inextricably related to the impossibility of separating the physical system under examination from the observer acquiring knowledge about the system. If we admit that measuring devices should be described by the same dynamical equations as the systems under consideration, then why does the measurement process break the superposition of states? This leads us to follow Bell \cite{Be04} in asking:  \\
\textit{``What exactly qualifies some physical systems to play the role of `measurer'?" } 

In this letter, we aim to provide a potential answer to this question. We postulate that physical systems act more or less as measuring devices depending on how much they exhibit a property called \textbf{quantum Integrated Information} (QII). This leads us to outline a novel, experimentally falsifiable theory with a universal dynamics depending on the levels of QII of physical systems. 

There have been numerous proposals to replace both unitary and measurement dynamics by a single, universal dynamics governing all physical processes \cite{Pea76,Ghi86,CSL90,Dio89,Pen96,Adl04}. Such a dynamical theory could be described using a non-linear, stochastic differential equation which does not allow superluminal signaling. We would then expect this equation to reduce to Schr$\ddot{o}$dinger's equation in the quantum regime and also provide an accurate description of the classical behavior of macroscopic objects. 

We stress that such a model aims to describe the physical world from an ontological perspective, whether or not any act of observation takes place. Knowledge about physical systems plays no fundamental role.  

An important question which naturally arises is the basis which should be chosen for the localization of the wavefunction. From our experience of macroscopic superpositions rapidly collapsing into localized states, it may seem that position should be considered as a privileged basis for collapse. We will discuss the role that relevant properties of the physical state could play in determining the basis on which the wavefunction is localized.  

From a phenomenological point of view, all space collapse models are equivalent: they induce a collapse of the wavefunction in space, such that the collapse rate depends on the size of the system. The assumption that the speed of localization of the system in space depends only on the size of the system but on none of its other properties seems rather ad hoc and naive. 

The key idea we explore here is that the relevant property of a physical system affecting the rate of collapse of the state might not be its size (or mass distribution) but should rather be related to its informational complexity. 

This naturally follows from the idea that quantum mechanical observers are expected to exhibit some form of `consciousness' which induces the wavefunction collapse. We take the view that consciousness plays a crucial role in quantum collapse and that conscious perceptions do not obey the linear laws of quantum mechanics. This leads to the difficult problem of finding a measure of consciousness. In the present work, we make no claims of having resolved this intricate philosophical issue but instead we take a working approach to this problem.

For the purpose of the present theory, we use a modified version of an existing `measure of consciousness', called Integrated Information (II) \cite{Ton04,Ton08}. The II of a physical system is defined as the information of the whole system above and beyond the information contained in its parts.

A quantum version of this measure (QII) enables us to explicitly present a novel \textit{Integrated Information-induced collapse theory}.

This theory may be interpreted as a modification of existing collapse models, where the rate of collapse of states is determined by a specific feature of their informational complexity: the QII \cite{Bas11}. We believe that this already provides an important conceptual shift, even if QII is completely unrelated to consciousness. 

This letter will first introduce the quantum Integrated Information, then will present the universal theory of Integrated Information collapse. We shall then describe potential experimental tests of the new theory in realms where it might not agree with quantum mechanics. Finally, we will discuss some of the modifications we might expect this collapse theory to undergo and sketch some issues that may arise.

\section{Consciousness and Integrated Information}

It has been suggested that physical systems exhibiting consciousness must satisfy two fundamental properties\cite{Baa88,Met03,Ton04,Ton08}. Firstly, \textit{differentiation of information} states that consciousness should allow discrimination of a single possibility amongst a vast repertoire of possible states, leading to the acquisition of information. Secondly, \textit{integration} is the feature that this differentiation should be performed by a unified physical system, not decomposable into a collection of independent parts. 

These concepts can be illustrated \cite{Ton04} by considering two unconscious physical systems. On the one hand, a digital camera with a million photodiodes exhibits a high level of differentiation but very little integration since it can enter a large number of distinct states but each photodiode acts independently. On the other hand, a million Christmas lights connected to a single switch exhibit a large amount of integration but almost no differentiation since either all the lights are on or they are all off. 
Both of these examples are in contrast with neural networks associated with consciousness in the human brain, since such physical systems are known to exhibit high levels of both differentiation of information and integration \cite{Mas07, Bal08}.

This observation hints that the amount of `consciousness' a physical system may manifest can be related to how much it exhibits a property called Integrated Information\cite{Ton04,Ton08}.     

In the present article, we define quantum Integrated Information (QII) as a general property of a quantum system, which corresponds to how much information the parts of a physical system contain above and beyond the information generated by the system as a whole. Therefore QII embodies this particular definition of consciousness as the capacity to process information in an integrated way.

\underline{Definition:} Given a quantum system in a Hilbert space $\mathcal{H}$ described by a density matrix $\rho$, we define the system's quantum Integrated Information as:
\begin{equation}\label{QII}
\Phi(\rho)=inf(S(\rho \vert \vert \bigotimes_{i=1}^{N} Tr_{\bar{i}}(\rho)): \mathcal{H} \cong_{\phi} \mathcal{H}_1 \otimes ... \otimes \mathcal{H}_N )
\end{equation} 

where we take the infimum over decompositions of the Hilbert space into subsystem Hilbert spaces $\mathcal{H}_i$ (by the isomorphism $\phi$).
The trace over $\bar{i}$ denotes the trace taken over all the subspaces other than the i subspace. Following the terminology of \cite{Ton08} we call the Hilbert space partition which minimizes the QII the minimum information partition (MIP). 
 
S is the quantum relative entropy:
\begin{equation}
S(\sigma_1 \vert \vert \sigma_2 ):= Tr(\sigma_1 log(\sigma_1))-Tr(\sigma_1 log(\sigma_2))
\end{equation} 
between the state of the system and the tensor product of the states obtained by tracing out each subsystem i in the MIP. 

Note that we can extend this definition to the case where the Hilbert space is decomposed into an infinite number of subspaces such that: $\mathcal{H} \cong \bigotimes_{i\in I} \mathcal{H}_i $, where the index set I is no longer the finite set \{1, ..., N \}.

An interesting question is whether the MIP always splits the Hilbert space into two subsystems. We expect that finding the MIP and calculating the QII of realistic physical systems will rely on the use of approximations and numerical techniques.

\section{Calculating the Quantum Integrated Information} 

We will now explicitly calculate the QII of two simple tripartite systems: the GHZ \cite{GHZ07} and W \cite{Du00} states.

The density matrices for these pure states are:
\begin{equation}
GHZ=\frac{1}{2}(\left|000\right\rangle+\left|111\right\rangle)(\left\langle 000\right|+\left\langle 111 \right|)
\end{equation}
\begin{equation}
W=\frac{1}{3}(\left|001\right\rangle+\left|010\right\rangle+ \left|100\right\rangle)(\left\langle 001\right|+\left\langle 010 \right|+\left\langle 100 \right|)
\end{equation}

Since both of these states are symmetrical, we only need to consider two candidate splittings for the MIP, namely separating the Hilbert space into three subsystems A, B and C or into two subsystems A and BC. Calculating the relevant reduced density matrices yields:
\footnotesize
\begin{equation}
GHZ_A \otimes GHZ_{BC}= \frac{1}{4} \left( \begin{array}{cccccccc}
1 & 0 & 0 & 0 & 0 & 0 & 0 & 0 \\
0 & 0 & 0 & 0 & 0 & 0 & 0 & 0 \\
0 & 0 & 0 & 0 & 0 & 0 & 0 & 0 \\
0 & 0 & 0 & 1 & 0 & 0 & 0 & 0\\
0 & 0 & 0 & 0 & 1 & 0 & 0 & 0\\
0 & 0 & 0 & 0 & 0 & 0 & 0 & 0 \\
0 & 0 & 0 & 0 & 0 & 0 & 0 & 0 \\
0 & 0 & 0 & 0 & 0 & 0 & 0 & 1 \end{array} \right)
\end{equation}
\begin{equation}
GHZ_A \otimes GHZ_{B} \otimes GHZ_{C}=\frac{\mathbb{I}}{8}
\end{equation}
\begin{equation}
W_A \otimes W_{BC}=\frac{1}{9}\left( \begin{array}{cccccccc}
2 & 0 & 0 & 0 & 0 & 0 & 0 & 0 \\
0 & 2 & 2 & 0 & 0 & 0 & 0 & 0 \\
0 & 2 & 2 & 0 & 0 & 0 & 0 & 0 \\
0 & 0 & 0 & 0 & 0 & 0 & 0 & 0\\
0 & 0 & 0 & 0 & 1 & 0 & 0 & 0\\
0 & 0 & 0 & 0 & 0 & 1 & 1 & 0 \\
0 & 0 & 0 & 0 & 0 & 1 & 1 & 0 \\
0 & 0 & 0 & 0 & 0 & 0 & 0 & 0 \end{array} \right)
\end{equation}
\begin{equation}
W_A \otimes W_{B} \otimes W_{C}=\frac{1}{27}\left( \begin{array}{cccccccc}
8 & 0 & 0 & 0 & 0 & 0 & 0 & 0 \\
0 & 4 & 0 & 0 & 0 & 0 & 0 & 0 \\
0 & 0 & 4 & 0 & 0 & 0 & 0 & 0 \\
0 & 0 & 0 & 2 & 0 & 0 & 0 & 0\\
0 & 0 & 0 & 0 & 4 & 0 & 0 & 0\\
0 & 0 & 0 & 0 & 0 & 2 & 0 & 0 \\
0 & 0 & 0 & 0 & 0 & 0 & 2 & 0 \\
0 & 0 & 0 & 0 & 0 & 0 & 0 & 1 \end{array} \right)
\end{equation}
\normalsize

Matrix diagonalization gives us: 
\begin{equation}
\log{(GHZ)}=\log{(W)}=\textbf{0}
\end{equation}

Therefore:

\footnotesize
\begin{equation}
\begin{split}
S(GHZ \vert \vert GHZ_A \otimes GHZ_{BC})&= -Tr(GHZ \log{(GHZ_A \otimes GHZ_{BC})}) \\
&= Tr \left( \begin{array}{cccccccc}
1 & 0 & 0 & 0 & 0 & 0 & 0 & 1 \\
0 & 0 & 0 & 0 & 0 & 0 & 0 & 0 \\
0 & 0 & 0 & 0 & 0 & 0 & 0 & 0 \\
0 & 0 & 0 & 0 & 0 & 0 & 0 & 0\\
0 & 0 & 0 & 0 & 0 & 0 & 0 & 0\\
0 & 0 & 0 & 0 & 0 & 0 & 0 & 0 \\
0 & 0 & 0 & 0 & 0 & 0 & 0 & 0 \\
1 & 0 & 0 & 0 & 0 & 0 & 0 & 1 \end{array} \right)=2
\end{split}
\end{equation}
\begin{equation}
S(GHZ \vert \vert GHZ_A \otimes GHZ_{B} \otimes GHZ_{C})=3
\end{equation}
\begin{equation}
S(GHZ \vert \vert W_A \otimes W_{BC})=2 \log{(\frac{3}{2})}\approx 1.17
\end{equation}
\begin{equation}
S(GHZ \vert \vert W_A \otimes W_{B} \otimes W_{C})=\frac{1}{3}(\log(\frac{27}{2})+2 \log(\frac{27}{4}))\approx 3.09
\end{equation}
\normalsize

Hence, we get that the QII of these states are: $\Phi($GHZ$)=2$ and $\Phi($W$)=2 \log{(\frac{3}{2})}\approx 1.17$.

\section{Integrated Information and state-vector reduction}

Quantum mechanics admits a clash between the linear deterministic evolution of an unobserved system and the nonlinear stochastic collapse of observed systems \cite{vN32,Bass00}. This dichotomy is at the heart of the difficulty in interpreting quantum theory and leads to the impossibility of attributing definite properties to physical systems independently of measurement. 

Collapse theories are alternatives to standard quantum mechanics, which aim to resolve these issues by presenting a universal non deterministic, nonlinear evolution law such that microprocesses and macroprocesses are governed by a single dynamics \cite{Pea76,Ghi86,CSL90,Dio89,Pen96,Adl04}.

We expect a universal dynamical equation to satisfy the following constraints, which strongly restrict the allowed form of the non-linear modification to Schr$\ddot{o}$dinger's equation:\\
(i) It must be almost identical to Schr$\ddot{o}$dinger's equation in the quantum regime but should break the superposition principle at the macroscopic level.  \\
(ii) It must be stochastic and should explain why measurement situations yield results distributed according to the Born rule. \\
(iii) It must not allow for superluminal signaling \cite{Gis89} in order to preserve relativistic causal structure. \\

Previous work on collapse models (see \cite{Bass13} for a review) has shown that a universal equation of the form:
\begin{equation}\label{gencollapse}
\frac{d}{dt}\rho(t)=-\frac{i}{\hbar}[H, \rho(t)] - \mathcal{I}[ \rho(t)]
\end{equation} 
where $\mathcal{I}$ is a non-linear operator representing the effect of the spontaneous collapse, can satisfy all three constraints.  

Standard space collapse models are astutely set up such that each particle undergoes random collapse leading to larger systems collapsing faster than small systems. In the dynamical equations, the rate of collapse is directly dependent on the number of particles or size of the physical system under study.

In our model, however, particles no longer undergo random collapse at random times but instead we consider that the spontaneous collapse follows from a type of group behavior. We expect that a physical system exhibiting a certain amount of informational complexity has an increased chance of spontaneous collapse. In that sense, we expect collapse to be less random than in other space collapse models: physical systems which have a high QII should naturally collapse faster.
 
We believe that a physical system's capacity to act as an observer should not depend on its size but on other physical properties instead. Indeed, localization follows from the process of observation which occurs in a measurement. This observation process taking place should require the observer in question to exhibit consciousness. This leads us to postulate that the main physical property determining whether or not a system can act as an observer is directly related to a key aspect of its informational complexity, namely its capacity to process information in an integrated way. 

The idea that a physical description of consciousness could be at the heart of resolving fundamental issues in quantum theory is not new \cite{Wig62,Sta11}. In the present article we make no claims of presenting such a description, but assume that quantum Integrated Information determines how much a system acts like an observer and exhibits spontaneous collapse.

We introduce a novel collapse model where the rate of collapse does not depend on a system's size but on how much QII it exhibits. The general evolution equation we propose is of the form:
\small
\begin{equation}\label{Limbcollapse}
\begin{split}
&\frac{d}{dt}\rho(t)=-\frac{i}{\hbar}[H, \rho(t)] \\
&+ \sum_{n,m=1}^{N^2-1} h_{n,m}(\Phi(\rho(t)))(L_n \rho(t)L_m^{\dagger}-\frac{1}{2}(\rho(t)L_m^{\dagger} L_n + L_m^{\dagger}L_n \rho(t))) 
\end{split}
\end{equation}
\normalsize

where the Hermitian matrix elements $h_{n,m}$ are continuous functions of the QII of $\rho$ (which are all zero when $\Phi(\rho)=0$) and $\{L_k \}$ is a basis of operators on the N dimensional system Hilbert space, which determines the basis in which the state collapses. 

If we assume Markovian behavior and no superluminal signaling then one can show \cite{Bad13} that this is the most general non-linear QII collapse equation. 
It has been argued \cite{Coh97,Bas03} that macro-objectification must take place in space and time and that position must therefore play the preferred role in collapse theories. Since space collapse models appear to be the only ones which explain the classical behavior of macroscopic objects, we must 
choose the $\{L_k \}$ basis such that the wavefunction localizes in the position basis. 

Hence, our model's objective description of how macroscopic reality arises is rather similar to the one resulting from the standard space collapse theories \cite{Ghi86,CSL90}, but where the mechanism causing the collapse onto the position basis depends on the QII. An underlying equation for wave function dynamics, whose general form would resemble that of standard space collapse models \cite{Bad13} but with parameters related to QII, could also provide an alternative description of our model.  

We can produce a large class of Integrated Information collapse models by replacing this evolution equation by equation (\ref{gencollapse}), with a more general non-linear operator $\mathcal{I}$ describing how the collapse rate depends on the system's QII. 

In the future, we expect a slightly modified version of the QII dynamical reduction equation to be compatible with relativity. This universal dynamics may emerge from a fundamental underlying theory in the spirit of trace dynamics \cite{Ad04,Ad13} or of quantum theory without spacetime \cite{Loch12}.

It could also turn out that the level of QII of a physical system is not the optimal measure of its capacity to encompass various distinguishable states and process information in a cohesive, integrated manner. Therefore, QII may have to be replaced by a more astute measure or one which is more convenient to calculate. We stress that the key idea of this article is that informational complexity, and more precisely the capacity to process information in an integrated manner, should replace size as the property of a physical system which determines its rate of collapse. Further details will require more fine tuning and input from experiments.

\section{Experimental tests of Integrated Information-induced collapse}

The Integrated Information collapse model we have presented here is an experimentally verifiable theory which is expected to yield some physical predictions which are in conflict with quantum mechanics. We will briefly discuss potential experiments which could serve to validate, reject or at least refine the new theory. 

The predictions of the new theory almost coincide with those of standard quantum mechanics at the microscopic level. Most current collapse models become significantly different from quantum theory when the size of the system under study increases. This leads to numerous experimental challenges due to the fact that environmental influences become more and more difficult to eliminate for larger systems. 

Typical experiments testing collapse models aim to set bounds on model parameters by studying the collapse of sizable physical systems in a large superposition \cite{Feld12, Nimm13, Arn14}. The aim of most superposition experiments is to observe spontaneous collapse of the wavefunction at a mesoscopic scale, after reducing the interaction with the environment. Tests of superposition include diffraction experiments with large molecules \cite{Arn99,Gerl11,Eil13},  optomechanical systems \cite{Marsh03}, microsphere interferometers \cite{Rom12} and indirect tests using cosmological data \cite{Ad07,Das13}.

Testing Integrated Information collapse is different from previous work on verifying the validity of collapse models. It is no longer sufficient to study large systems in order to increase the predicted rate of collapse. Indeed, we expect novel behavior in conflict with quantum theory to arise in situations where physical systems with a high level of QII exhibit non-linear collapse and cause a breakdown of the quantum principle of linear superposition.

Therefore, the first step in verifying QII collapse consists of calculating the quantum Integrated Information of various interesting physical systems. This may require some numerical approximations and clever optimization in order to determine the minimum information partition (MIP) for each system. 

The next step would then be to compare the collapse rate of various physical systems with very different QII. We expect these experiments testing quantum superposition to be similar in nature to current collapse model tests. They would require an extremely precise control of the environment since the effects of decoherence need to be accounted for to a high precision. Note that one would expect conscious beings to clearly exhibit high levels of QII and therefore physical systems including such beings would undergo spontaneous collapse. It may be the case, however, that certain complex inanimate objects may have a high QII and therefore also behave as observers, in the sense that their presence within a larger physical system leads to collapse.

In some respects, the experimental tests of QII collapse models may be simpler to implement than those for standard spontaneous collapse since the systems under examination might not have to be as large. Indeed, several relatively small mesoscopic systems of similar size may exhibit very different levels of QII and have observably different spontaneous collapse rates. 

These experiments should help us refine the collapse model dynamics and determine the $h_{n,m}(\Phi)$ matrix elements in equation (\ref{Limbcollapse}). They will also lead to a better understanding of whether QII is indeed the best measure of a physical system's capacity to spontaneously collapse. 

\section{Conclusion}

We have presented a novel theory which is in conflict with quantum mechanics. Even if it turns out that QII spontaneous collapse does not agree with future experiments, we feel that the theoretical implications of the new collapse theory are of interest for their own sake and may shed some light on various features of quantum theory.

First of all, it may be interesting to study computational properties of the new collapse model. How would the spontaneous collapse of systems with high QII affect the possibility of performing large `quantum' computations. Can one define a modified version of many-worlds theory which can be related to the QII collapse model?

Moreover, we believe that the basis on which wavefunction localization takes place should not always be position. The relationship between another physical definition of Integrated Information and the so-called \textit{quantum factorization problem} has been addressed in \cite{Teg14}. 

In general, we expect that the collapse basis for each system may depend on properties of a quantum version of qualia space \cite{Bald09}, corresponding to the quality of consciousness of the system in question. In this sense, dynamics would not just be governed by the QII of a physical system but also by the set of all the informational relationships that causally link its elements.

In the model we are currently proposing, the collapse mechanism is universal and not related to specific systems since position plays a fundamental role, similarly to the current spontaneous collapse models. Further work, however, could redefine equation (\ref{gencollapse}) and the operator basis $\{L_k \}$ such that the collapse basis is different for each physical system in a way which explains the apparent fundamental role of the position basis. Space-time would then emerge from the fact that we cannot extract ourselves from the physical systems we examine.

This may lead to alternative versions of quantum field theory, where space-time does not play a fundamental role. We expect new particles -- \textit{complexetrons}-- to arise due to the spontaneous collapse term in equation (\ref{gencollapse}).

We look forward to revealing the physical world described by Integrated Information-induced collapse.

\bibliography{Conscbib}

\end{document}